\documentclass[a4paper,fleqn,usenatbib,useAMS]{mnras}

\usepackage[T1]{fontenc}
\usepackage{ae,aecompl}

\usepackage{graphicx}	
\usepackage{amsmath}	
\usepackage{amssymb}	

\title[Could multiple voids explain the CMB Cold Spot?]{Could multiple voids explain the Cosmic Microwave Background Cold Spot anomaly?}

\author[Krishna Naidoo, Aur\'{e}lien Benoit-L\'{e}vy \& Ofer Lahav]{
Krishna Naidoo$^{1}$\thanks{E-mail: \href{mailto:krishna.naidoo.11@ucl.ac.uk}{krishna.naidoo.11@ucl.ac.uk}}, Aur\'{e}lien Benoit-L\'{e}vy$^{2,3,4,1}$\thanks{E-mail: \href{mailto:benoitl@iap.fr}{benoitl@iap.fr}} and Ofer Lahav$^{1}$\thanks{E-mail: \href{mailto:o.lahav@ucl.ac.uk}{o.lahav@ucl.ac.uk}}
\\
$^{1}$Department of Physics \& Astronomy, University College London, Gower Street, London WC1E 6BT, UK\\
$^{2}$Sorbonne Universit{\'e}s, UPMC Univ Paris 6 et CNRS, UMR 7095, Institut d'Astrophysique de Paris,\\
98 bis bd Arago,75014 Paris, France\\
$^{3}$Fermi National Accelerator Laboratory, P.O. Box 500, Batavia, IL 60510, USA\\
$^{4}$Kavli Institute for Cosmological Physics, University of Chicago, Chicago, IL 60637, USA\\
}

\date{Accepted XXX. Received YYY; in original form ZZZ}

\pubyear{2016}

\begin{document}
\label{firstpage}
\pagerange{\pageref{firstpage}--\pageref{lastpage}}
\maketitle

\begin{abstract}
Understanding the observed Cold Spot (CS) (temperature of $\sim -150 \mu K$ at its centre) on the Cosmic Microwave Background (CMB) is an outstanding problem. Explanations vary from assuming it is just a $\gtrsim3\sigma$ primordial Gaussian fluctuation to the imprint of a supervoid via the Integrated Sachs-Wolfe and Rees-Sciama (ISW$+$RS) effects. Since single spherical supervoids cannot account for the full profile, the ISW$+$RS of multiple line-of-sight voids is studied here to mimic the structure of the cosmic web. Two structure configurations are considered. The first, through simulations of 20 voids, produces a central mean temperature of $\sim-50\mu K$. In this model the central CS temperature lies at $\sim2\sigma$ but fails to explain the CS hot ring. An alternative multi-void model (using more pronounced compensated voids) produces much smaller temperature profiles, but contains a prominent hot ring. Arrangements containing closely placed voids at low redshift are found to be particularly well suited to produce CS-like profiles. We then measure the significance of the CS if CS-like profiles (which are fitted to the ISW$+$RS of multi-void scenarios) are removed. The CS tension with the $\Lambda$CDM model can be reduced dramatically for an array of temperature profiles smaller than the CS itself.
\end{abstract}





\section{Introduction}

The Cosmic Microwave Background (CMB) has proven to be one of the most powerful tools for precision Cosmology. The CMB is in remarkable agreement with theoretical predictions, but there remain certain anomalies that are not fully understood. One such anomaly is the CMB Cold Spot (CS), which was discovered \citep{Vielva2004} when searching for non-Gaussianity in WMAP CMB maps. This was done using the Spherical Mexican Hat Wavelet (SMHW) \citep{Cayon2001} and was only observed in 0.1\% of 10,000 Gaussian simulations \citep{Cruz2005}. More recently, the CS hot ring has been shown to be its defining anomalous feature \citep{Zhang2010,Nadathur2014,PlanckIso2015}. The CS has been studied in great detail \citep{Cruz2006, Wmap2007, Cruz2008, Wmap2011, Gurzadyan2014, PlanckIso2015} and while many models have been put forward to explain its origin, observational evidence to verify them have been lacking.

In recent years, one of the leading emergent models states the CS is the imprint of a large void \citep{Inoue12007,Inoue22007,Masina2009,Inoue2010,Inoue2012}. Light passing through the void will experience a change in energy due to the time evolution of gravitational potentials via the Integrated Sachs-Wolfe (ISW) \citep{Sachs1967} and Rees-Sciama (RS) \citep{Rees1968} effects. Momentum for this idea has been fueled by cross-correlation studies of the CMB with large-scale structure (LSS) which, in some cases, have found higher than predicted correlation \citep{Rassat2007,Giannantonio2008, Granett2008, Papai2011, Nadathur2012, Flender2013, Manzotti2014, Andras2015}. Up until recently, searches for voids along the CS line-of-sight (LOS) have failed to find any significant underdensities \citep{Smith2010,Granett2010,Manzotti2014}, in particular pencil beam surveys \citep{Granett2010, Bremer2010} have effectively ruled out underdensities in the CS LOS between $0.3<z<1$. A large supervoid (the so called Eridanus Void) was finally discovered by \citet{Szapudi2015}, but \citet{Nadathur2014} (NLHR) and \citet{Zibin2014} showed single spherical structures could not explain the CS profile.

Despite the numerous studies on the possible ISW and RS (ISW$+$RS) imprints of single voids to produce the CS, only a few studies \citep[such as][]{Finelli2015} have attempted to look at the effect of multiple structures along the LOS (which would imprint a larger than expected temperature profile). The large scale cosmic web structure \citep{Bond1996} of the Universe, makes the alignment of -- at the very least -- a few voids seem very plausible. Hence, toy models of multiple voids are tested. Although the equations used in this study assume spherical symmetry, tightly stacked voids can be considered as representative of elongated voids \citep[explored in][]{Marcos2015,AndrasJuan2015}. These have been found to be the possible cause of larger than expected cross-correlations between LSS and the CMB \citep{Granett2015}.

\section{Modeling}
\label{ISWRStheory}

For a spherical void the radius ($r_{0}$), \emph{matter} density contrast depth ($\delta_{0}$) (which has the opposite sign convention to density contrast), and central redshift ($z_{0}$) are required to calculate the corresponding expected ISW$+$RS imprint. For the Eridanus void, these parameters are: $r_0 = 220 \pm 50 h^{-1} Mpc$, $z_{0} =0.22\pm0.03$ and $\delta _{0} = 0.38 \pm 0.11$\footnote{Obtained from $\delta_{0}=-\delta_{m}e^{1}$ (NLHR) where $\delta_{m}=-0.14 \pm 0.04$ \citep{Szapudi2015} is the average \emph{matter} density contrast across the void profile assuming a biasing factor of $b=1.41$.} \citep{Szapudi2015} with $1\sigma$ errors.

In this study the density contrast profile of spherical voids \citep[introduced by][]{Finelli2015} are defined by equation~\ref{eq:delta}. Two void models are examined: the $\alpha=0$ case -- considered by NLHR -- and the $\alpha=1$ -- an alternative model introduced by \citet{Finelli2015}. The $\alpha=1$ case contains a more pronounced compensated region which allows for a hot ring feature to be produced via the ISW effect (which dominates in both cases),
\begin{equation}
\label{eq:delta}
\delta (r) = -\delta _{0} \left( 1 - \frac{2+7\alpha}{3+3\alpha} \frac{r^{2}}{r_{0} ^{2}} + \frac{2\alpha}{3+3\alpha} \frac{r^{4}}{r_{0} ^{4}} \right) \exp \left(- \frac{r^{2}}{r_{0} ^{2}} \right).
\end{equation}

For $\alpha=0$ we employ the model derived by NLHR, while for $\alpha=1$ we use \citet{Finelli2015}\footnote{This study began prior to the latest release of \citet{Finelli2015}, hence the $\alpha=0$ case was performed using solely the NLHR approach. However, consistent results for $\alpha=0$ can be obtained from \citet{Finelli2015}.}. The ISW$+$RS effect is then computed by adding the ISW and RS effects across the projected angular distance ($\theta$) on the sky, 
\begin{equation}
\label{eq:ISWRS}
\Delta T _{\rm ISW+RS} (\theta) = \Delta T_{\rm ISW} (\theta)+ \Delta T_{\rm RS} (\theta).
\end{equation}

A toy model approach is utilised to calculate the imprint of multiple voids. Simulations are run using void catalogues, which contain the void's effective radius ($R_{eff}$) and $z_{0}$. However, to calculate $\delta_{0}$, the central normalised galaxy count $(\rho/\bar{\rho})_{g,0}$ is used. $\delta _{0}$ is then calculated from $\delta _{0} = -(1/b)[(\rho/\bar{\rho})_{g,0}-1]$ using a linear bias of $b=1.9$ \citep{Max2006}. For simplicity we assume $R_{eff}$ is equivalent to $r_{0}$. Although in practice it is unlikely that the structures will match the density contrast profiles assumed. Even in the case that they do, it is improbable that $r_{0}$ would be precisely picked up by the void finder algorithms. Nevertheless, this provides us with an indicative ensemble of voids. The ISW$+$RS temperature profiles of multiple structures is then found by superimposing the imprints of single structures along the LOS\footnote{Note, the ISW$+$RS of many voids placed close together is non-linear for the RS effect (for $\delta_{0} \gtrsim 0.5$), so the process is not completely additive as is assumed.}.

A $\Lambda$CDM model is assumed with cosmological parameters: $\Omega _{m} =0.27$, $\Omega _{\Lambda} = 0.73$, $h = 0.7$ and $T_{CMB} = 2.7255K$. Although these slightly differ from \citet{PlanckPara2015}, they are adopted for ease of comparison with NLHR. Note, the density contrast profile of these structures extends further out than $r_{0}$. When structures are placed together, the extended regions (beyond $r_{0}$) overlap. In reality such a superposition is not possible, but the temperature profiles generated are indicative of what the true profile would be.

\section{N Eridanus Supervoids}

\begin{figure}
	\centering
	\includegraphics[width=\columnwidth]{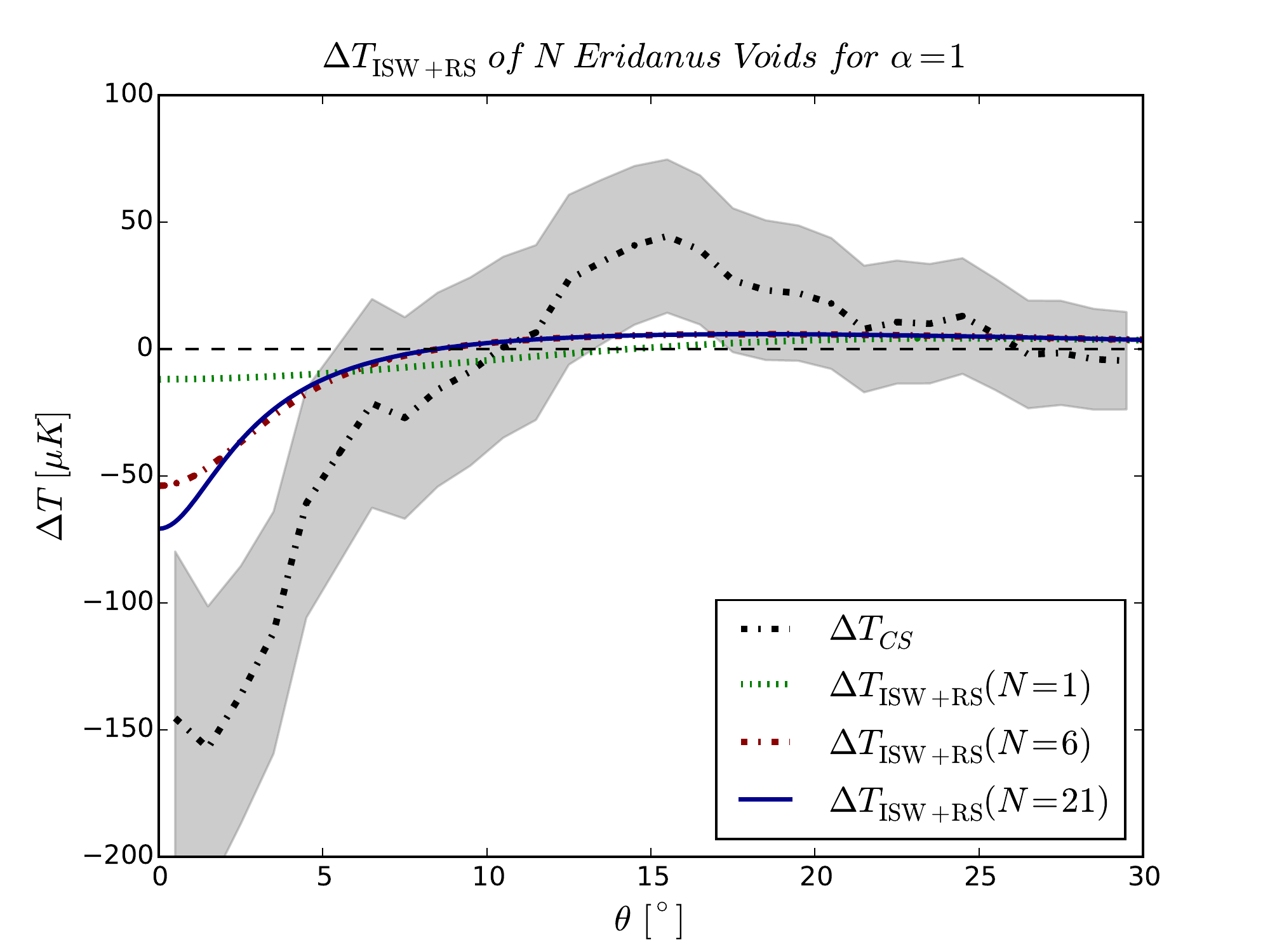}
	\caption{The $\alpha=1$ ISW$+$RS of 1, 6 and 21 Eridanus voids, placed end to end along the LOS, is plotted with the CS temperature profile (dashed-dotted line) and its associated standard deviation (gray contour). The central CS temperature is not matched even with the maximum 21 voids, however the addition of Eridanus voids does make the profile more CS-like.}
	\label{fig:N2}
\end{figure}

The temperature profile of the CS (see black dashed-dotted line in Figures~\ref{fig:N2} and \ref{fig:ISWRS_V_lrg}) was remeasured in this study using the Planck SMICA map (downgraded to $N_{\rm side}=256$) by measuring the average temperature in concentric rings, of $1^{\circ}$ radius, from the CS centre (located at galactic coordinates $(l,b) \sim (209^{\circ},-57^{\circ})$). 

To begin, the effects of N Eridanus voids are computed by placing them end to end along the LOS. The maximum number of stacked Eridanus voids along the LOS is 21. This is calculated by dividing the comoving distance to last scattering by the diameter of the Eridanus void. The first Eridanus void is placed according to its true redshift centre and the centre of each subsequent Eridanus void is then placed $2\times r_{0}$ (in comoving coordinates) from the centre of the previous void. Simulations involving many Eridanus voids become increasingly unrealistic; as these place voids at higher redshift where: (1) the matter distribution is not sufficiently clumpy for large underdensities to be present and (2) no meaningful ISW signal should be expected since dark energy would not be as dominant. For $\alpha=0$, the central CS temperature can be matched by 6 Eridanus voids stacked along the LOS. However, this does not feature a hot ring, which makes it an incompatible CS explanation. In Figure~\ref{fig:N2} the $\alpha=1$ temperature profiles are computed for 1, 6 and 21 Eridanus voids. Even with the maximum number, the central CS temperature cannot be achieved. However, the profile quickly moves towards one that is similar in shape (up to $10^{\circ}$). Multiple Eridanus void arrangements are clearly favoured, but this string of stacked underdensities considered may itself be more problematic for $\Lambda$CDM than the CS itself.

\section{N lrg Voids}

The temperature profile of multiple voids along the LOS is calculated by drawing void parameters from LSS catalogues. The motivation is to calculate the ISW$+$RS profile of an ensemble of voids that are representative of the distribution of voids in the Universe.

\subsection{Data}

To calculate the overall effect of multiple LOS voids, we use the publicly available cluster and void catalogue\footnote{Available from \url{http://research.hip.fi/user/nadathur/download/dr7catalogue/}, where the latest version includes data from SDSS Data Release 11.} compiled by \citet{NadathurHotchkiss2014}. This uses spectroscopic data from Data Release 7 of the Sloan Digital Sky Survey. The \emph{lrgbright} and \emph{lrgdim} catalogues of \emph{Void Type 1} and \emph{Type 2} are combined. This catalogue (referred to as the \emph{lrg} voids) consists of 103 voids. These voids are generally quite large since they are found using sparser galaxy tracers.

\subsection{Method}

1,000 simulations for the ISW$+$RS effects of 10 and 20 voids were run for $\alpha=0$, while 10,000 were run for $\alpha=1$. More simulations are carried out for $\alpha=1$ since these simulations produce hot rings. In these simulations, void parameters $\delta_{0}$ and $r_{0}$ are randomly drawn from our \emph{lrg} catalogue and placed at random between redshift $z=0.1-2$, whilst ensuring that the voids do not overlap. The lower bound of $z=0.1$ is chosen as a buffer so that the observer is not mistakenly placed within the structure which would require the inclusion of a dipole effect \citep{Nadathur2014, Zibin2014}. For each value of $\theta$ the temperature values for all simulations are taken and sorted in ascending order. 1$\sigma$ and 2$\sigma$ are then found by ensuring 68\% and 95\% of the data lies within these envelopes respectively.

\begin{figure}
	\centering
	\includegraphics[width=\columnwidth]{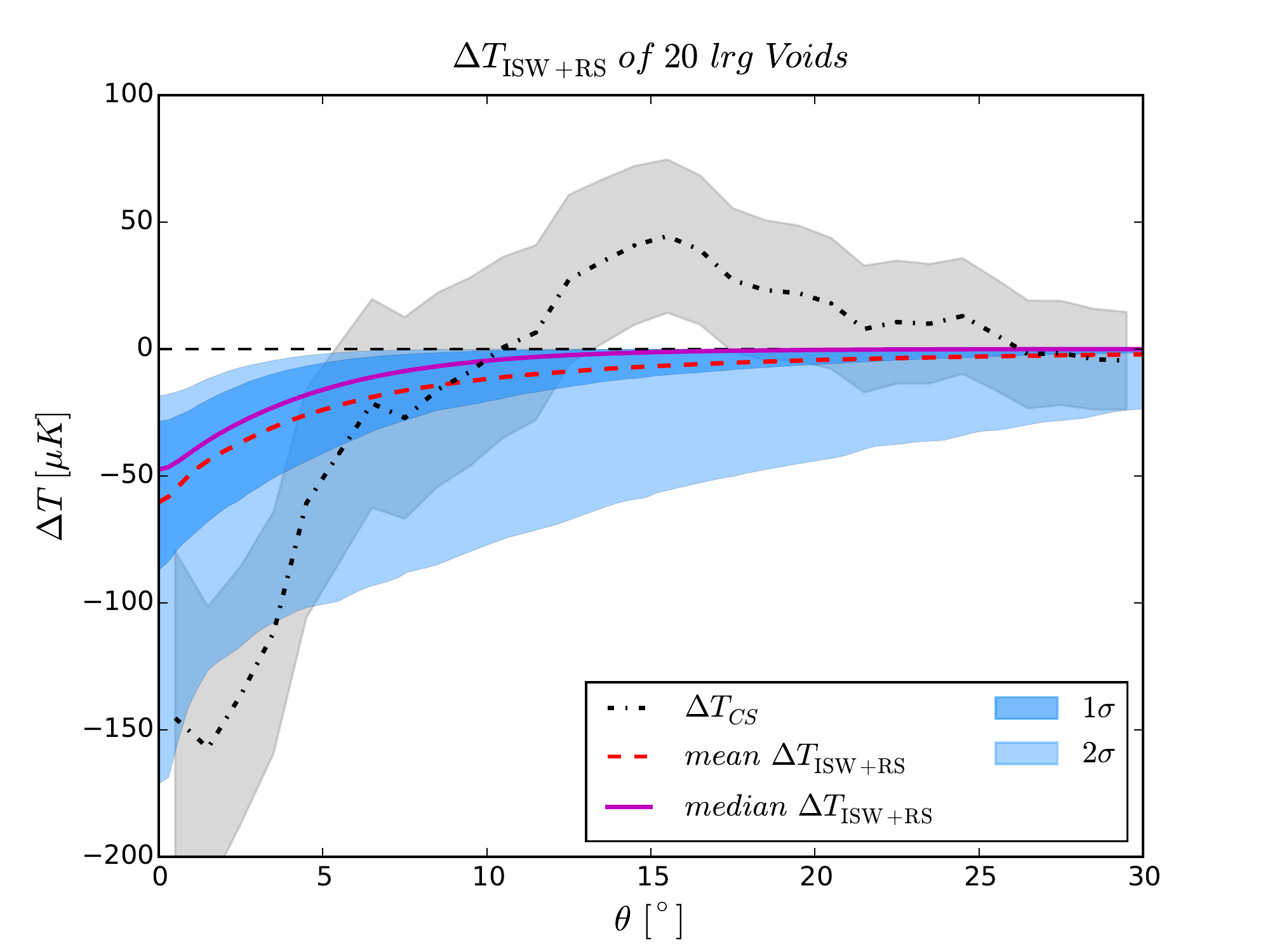} 
	\caption{The mean, median and general $\alpha=0$ ISW$+$RS profile of $N_{\rm{v}}=20$ lrg voids are plotted in comparison to the CS temperature profile (dashed-dotted line) with its associated standard deviation (gray contour).}
	\label{fig:ISWRS_V_lrg}
\end{figure}

\begin{figure*}
\centering
\includegraphics[width=1\textwidth]{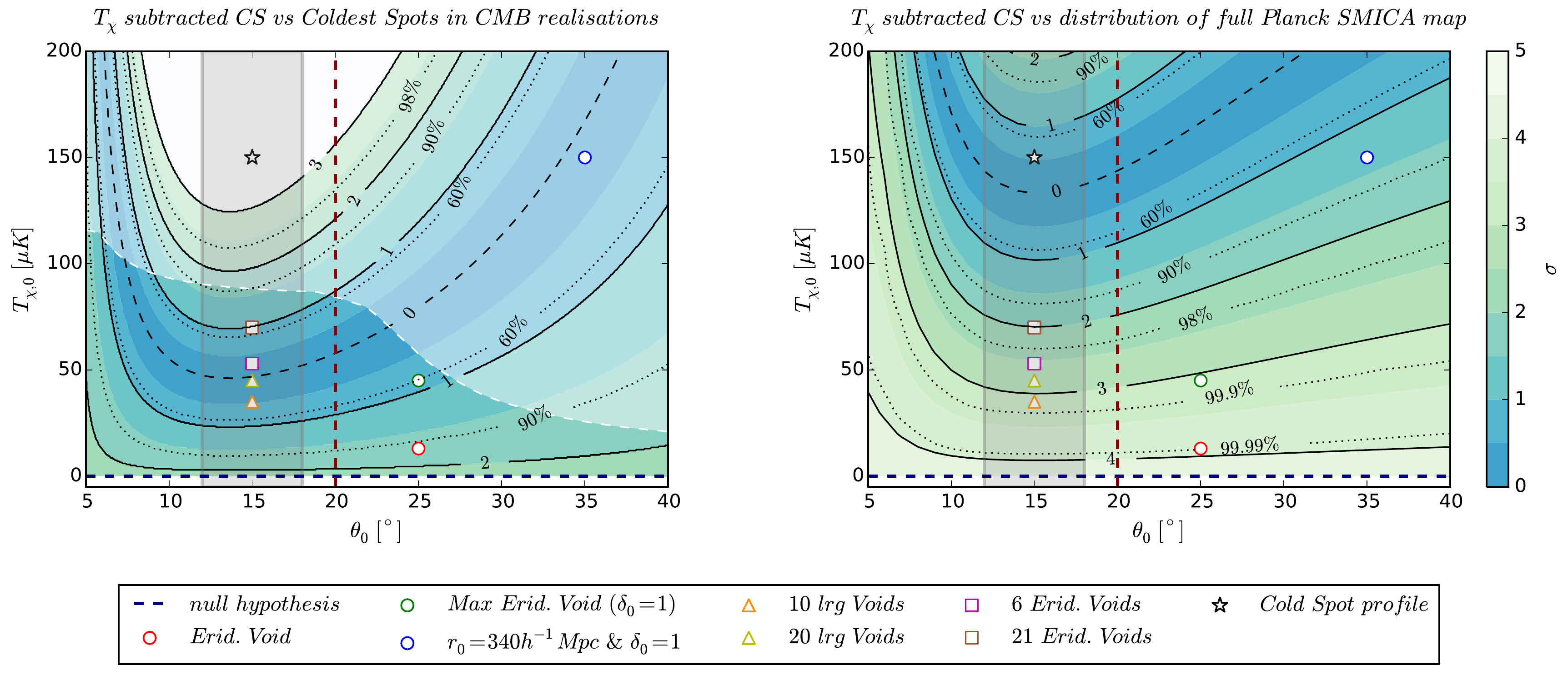}
\caption{On the \emph{left} the $T_{\chi}$ subtracted CS (or residual temperature profile) is compared to the coldest spots in CMB realisations, following the procedure carried out in Figure 7 of NLHR. On the \emph{right}, the SMHW of the residual temperature profile is compared to the full SMHW filtered SMICA map (at $N_{\rm side}=16$). The white shaded region shows where the residual temperature profile deviates by more than $2\sigma$ from the mean coldest spots in CMB realisations (identified using procedure shown in NLHR). The CS significance under the null hypothesis (a completely Gaussian primordial fluctuation) is shown by the dashed blue line. While the significance of the CS if the following profiles are \emph{removed} is indicated by: (1) CS (black star), (2) Eridanus void (red circle), (3) Eridanus void with $\delta_{0}=1$ (green circle), (4) large void that can match the central CS temperature (blue circle), (5) the deepest profiles for CS candidates found in simulations with $10$ (orange triangle) and $20$ (yellow triangle) \emph{lrg} voids, and (6) 6 (pink square) and 21 (brown square) Eridanus voids. Note, only the ISW$+$RS temperature imprints of $\alpha=1$ are shown. These profiles are fitted approximately to the $T_{\chi}$ parameters. The region beyond $\theta_0>20^{\circ}$ (red dashed line), marks the angle from which the SMHW filter (used in either measurement) approaches 0 and the position of the CS hot ring is marked by the grey shaded region between $12-18^{\circ}$.}
\label{fig:CS_significance}
\end{figure*}
\subsection{ISW$+$RS for $\alpha=0$}
\label{sec:a=0}

10 and 20 voids (where the number of voids is denoted by $N_{\rm{v}}$) are placed along the LOS between redshifts $0.1$ and $2$. While our study did consider scenarios involving both voids and clusters, these arrangements were found to be of no particular interest in regards to the CS as these profiles are too shallow. We focus instead on the results produced using only \emph{lrg} voids. In Figure~\ref{fig:ISWRS_V_lrg} the ISW$+$RS temperature profile distribution for $N_{\rm{v}}=20$ is shown exhibiting a central mean temperature of $\sim -50 \mu K$. However, the deepest profiles are capable of explaining the central CS region (as it is shown to lie at $\sim 2\sigma$), but cannot produce the hot ring. This makes it difficult to consider this as a viable explanation for the CS.

\subsection{ISW$+$RS for $\alpha=1$}

Introduced by \citet{Finelli2015}, this void profile contains a more pronounced compensated shell, ensuring that a hot ring is produced via the ISW effect. The ISW$+$RS profiles of $N_{\rm{v}}=10$ and $20$ voids were simulated and are much shallower than the $\alpha=0$ model and is therefore unlikely to explain the full CS profile. However, perhaps only a fraction of the full CS profile is required \citep{Inoue2012} (see section~\ref{sec:sig}). To search for CS-like profiles (or CS candidates), simulations are sorted (which place voids between $z=0.1-2$; in the same manner as in section~\ref{sec:a=0}) in search of ones with peaks between $\theta = 12-18 ^{\circ}$ - to find profiles which have a hot ring feature aligned with the CS. These profiles tend to occur for simulations where the distribution of voids is higher than average at low redshifts ($0.1-0.5$). Hot ring features are found in $11.26\%$ of the $N_{\rm{v}}=10$ simulations which increases to $13.73\%$ for $N_{\rm{v}}=20$ simulations. This indicates that tightly arranged voids (or an elongated void) at low redshift are favoured as CS candidates.

\section{How much of the Cold Spot needs to be explained?}
\label{sec:sig}

We examine the significance of the primordial CS feature if a fraction of the CS observed is caused by another effect, like the ISW$+$RS or some hitherto unknown effect. Temperature profiles lying between the null hypothesis (i.e. a completely primordial Gaussian fluctuation) and the full CS profile are subtracted. This is defined by equation~\ref{equ:T_chi}, to be similar in shape to the CS -- i.e. a central decrement followed by a hot ring. $T_{\chi,0}$ defines the depth of the profile and $\theta_{0}$ defines the angle at which the hot ring peaks. If we set $T_{\chi, 0} = 150 \mu K$ and $\theta_{0} = 15 ^{\circ}$, we acquire a profile which resembles the CS closely.
\begin{equation}
T_{\chi} ( \theta )= - T_{\chi,0} \left( 1 - \frac{12}{5} \Bigg( \frac{\theta}{\theta_{0}} \right) ^{2} \Bigg) \exp \Bigg( - \frac{17}{10} \Bigg( \frac{\theta}{\theta_{0}} \Bigg) ^{2} \Bigg)
\label{equ:T_chi}
\end{equation}

The values of $T_{\chi,0}$ and $\theta_{0}$ are varied to construct a $T_{\chi}$ profile which is then subtracted from the CMB SMICA map in the direction of the CS. The SMHW convolution is then applied to this $T_{\chi}$ subtracted map at $N_{\rm side}=16$, with an angular scale of $R=5^{\circ}$ \citep[following ][]{Zhang2010}. We first compare the centre of the CS in the $T_{\chi}$ subtracted maps (carried out at $N_{\rm side}=16$) to the overall distribution of the original CMB SMHW map (at $N_{\rm side}=64$) for a particular SMHW filter. This is shown in Figure~\ref{fig:CS_significance} (\emph{right}). The positions of several ISW$+$RS profiles are placed approximately, alongside the current $\alpha=1$ ISW$+$RS profile of the Eridanus void which is shown to do little to the significance of the CS. Interestingly, we find that the CS can be reduced to a $2\sigma$ primordial fluctuation if a temperature profile with an aligned hot ring and central temperature of $\sim-70\mu K$ are considered. However, only the most extreme profile (a void with $r_0=340$ $h^{-1} Mpc$ and $\delta_{0}=1$) actually reduces the significance to below $2\sigma$. The other profiles still leave the CS with a significance of $\gtrsim 2\sigma$, despite involving extreme scenarios. So while a smaller temperature profile could explain the CS we have shown that even this reduced temperature profile is difficult to achieve through the ISW$+$RS effect. 

We then compare the residual temperature profile (i.e. the $T_{\chi}$ subtracted CS) to the coldest spots in CMB realisations. This is carried out in the same manner as is shown in NLHR, where the significance is calculated by measuring the cumulative filtered temperature up to $20^{\circ}$ but with an angular scale $R=5^{\circ}$ \citep[see Figure 7 of ][]{Nadathur2014}. The values found are plotted on the \emph{left} of Figure~\ref{fig:CS_significance}, where a further constraint is applied by calculating the difference between the residual temperature profile and the mean coldest CMB spots. In Figure~\ref{fig:CS_significance} (\emph{left}) the region shaded in white, illustrates where the maximum significance of the residual profile deviates by more than $2\sigma$ from the mean temperature profile of the coldest CMB spots. The region not shaded in white with a significance of $<2\sigma$ are the types of profiles that would make the CS more consistent with the coldest CMB spots. A profile with $\theta_{0}\sim15^{\circ}$ and $\Delta T_{\chi,0}\sim-50\mu K$ would greatly reduce the anomaly. This is well within the scenarios considered, however voids beyond $z>0.3$ have been effectively ruled out.  Nevertheless the CS anomaly could arise due to a crucial ISW$+$RS imprint from the Eridanus void which stretches along the entire LOS region ($z<0.3$) \citep{AndrasJuan2015}, which is precisely the region where our simulations have found the best case for a causal relation.

\section{Conclusion}

To test whether stacked voids can produce the CS anomaly, simulations were run using two density contrast profiles (defined by $\alpha=0$ and $\alpha=1$) (see eq~\ref{eq:delta}). 
For $\alpha=0$, we find that the ISW$+$RS for 6 Eridanus voids and 20 \emph{lrg} voids is capable of providing the central CS temperature. However, this model fails to predict a hot ring, making it incompatible with the full CS profile. For $\alpha=1$ (which contains a more pronounced compensated region), we find that the ISW$+$RS cannot provide an explanation for the full CS amplitude for the void simulations considered. Nevertheless, arrangements with an excess of voids at low redshift tend to have aligned hot rings. Furthermore, stacking 6 and 21 Eridanus voids along the LOS reveals that the elongation of the N Eridanus void structure makes the temperature profile more CS like (although not as deep). In both cases, the temperature profiles that are most CS-like tend to be scenarios involving tightly arranged voids along the LOS, an arrangement that could be interpreted as a pseudo-elongated void.

A temperature profile, $T_{\chi}$ (see equation~\ref{equ:T_chi}), is defined with CS-like characteristics. The parameters are varied and then subtracted from the CMB SMICA map in the direction of the CS. The significance of the residual temperature profile is measured in comparison to the coldest spots in CMB realisations and the full SMICA map distribution. We find that a whole range of profiles could relieve the CS tension with $\Lambda$CDM and display that a temperature profile that is as deep as the CS is not necessarily required. A temperature imprint with an aligned hot ring and a central temperature of $\sim - 70 \mu K$ would be sufficient to reduce the CS significance to $2\sigma$. However, even this reduced temperature profile is shown to be difficult to achieve through the ISW$+$RS for $\alpha=1$. Alternatively, for the CS to be consistent with the coldest spots in CMB realisations requires a central $\Delta T_{\rm ISW+RS}\sim -50\mu K$. Multiple void structures at low redshift seem to suggest a stronger causal relation between the voids and CS. Since the Eridanus void is both elongated and at low redshift \citep{AndrasJuan2015}, a causal relation now seems very likely. Hence, it is crucial that future galaxy surveys map the cosmic web along the LOS to better understand the shape and role of this void and other structures in the CS anomaly.

\section{Acknowledgement}

We thank Seshadri Nadathur, Andr\'{a}s Kov\'{a}cs and Juan Garc\'{i}a-Bellido for correspondence and discussion. Some of the results in this paper were derived using the HEALPix \citep{Healpix2005} package. ABL thanks CNES for financial support through its post-doctoral programme. This work was supported in part by the Kavli Institute for Cosmological Physics at the University of Chicago through grant NSF PHY-1125897 and an endowment from the Kavli Foundation and its founder Fred Kavli. OL acknowledges support from a European Research Council Advanced Grant FP7/291329, which also supported ABL and KN.




\bibliographystyle{mnras}
\bibliography{bibfile} 




\label{lastpage}
\end{document}